# Laser-synthesized TiN nanoparticles as novel efficient sorbent for environmental water cleaning

**Alexander V. Syuy[1], Ilya V. Martynov[1], Ilya A. Zavidovskiy[1], Dmitry V. Dyubo[1], Qingjiang Sun[2], Xi Yang[2], Gleb V. Tikhonowski[3], Daniil I. Tselikov[1,3], Maxim S. Savinov[3], Islam V. Sozaev[3], Anton A. Popov[3], Sergey M. Klimentov[3], Gleb I. Tselikov[4], Valentyn S. Volkov[4], Sergey M. Novikov[1], Aleksey V. Arsenin[1,5], Xiangwei Zhao[2], Andrei V. Kabashin[6,\*]**

[1] Center for Photonics and 2D Materials, Moscow Institute of Physics and Technology, 141701 Dolgoprudny, Russia; alsyuy271@gmail.com (A.V.S.); martynov.iv@mipt.ru (I.V.M); zavidovskii.ia@mipt.ru (I.A.Z.); novikov-s@yandex.ru (S.M.N.); dyubo.dv@phystech.edu (D.V.D.); arsenin.av@mipt.ru (A.V.A.)

[2] State Key Laboratory of Digital Medical Engineering, School of Biological Science & Medical Engineering, Southeast University, Nanjing, 210096, China; xwzhao@seu.edu.cn (X.Z.); sunqj@seu.edu.cn (Q.S.); 101013370@seu.edu.cn (X.Y.)

[3] Laboratory "Bionanophotonics", Institute of Engineering Physics for Biomedicine (PhysBio), MEPhI, Moscow 115409, Russia; gvtikhonovskii@mephi.ru (G.V.T.); ditselikov@gmail.com (D.I.T.); savinov.maxim.just@gmail.com (M.S.S.); islam.sozaev@mail.ru (I.V.S.); aapopov1@mephi.ru (A.A.P.); smklimentov@mephi.ru (S.M.K.)

[4] Emerging Technologies Research Center, XPANCEO, Internet City, Emmay Tower, Dubai, United Arab Emirates; celikov@xpanceo.com (G.I.T.); vsv@xpanceo.com (V.S.V.)

[5] Laboratory of Advanced Functional Materials, Yerevan State University, Yerevan 0025, Armenia; arsenin.av@mipt.ru (A.V.A.)

[6] CNRS, LP3, Aix-Marseille Université, Marseille 13288, France; andrei.kabashin@yahoo.com (A.V.K.)

\* Correspondence: andrei.kabashin@yahoo.com ; Tel.: +33 7 60 60 68 74

**Abstract:** Dyes used in industries such as textile, paper, and leather are known to be harmful to both human health and aquatic ecosystems. Therefore, finding effective and sustainable methods to remove dyes from wastewater is crucial for mitigating the detrimental effects of pollution. TiN nanoparticles have good absorption and conversion of light energy into thermal energy in the visible range of the spectrum, which makes them promising in various applications, from biomedical to environmental protection. In this work, it is shown that titanium nitride nanoparticles also possess promising adsorption capabilitieseffect. TiN nanoparticles were synthesized by laser ablation method in liquid. Water, acetone and acetonitrile are used as solvent. Nanoparticles were characterized by scanning and transmission microscopy, Raman spectroscopy, which showed the formation of the under-stoichiometric titanium nitride ($TiN_{1-x}$). TiN nanoparticles are investigated as a promising object for high adsorption It is shown that adsorption of TiN nanoparticles is associated with the electrostatic effect and the presence of pores in the synthesized nanoparticles. Optimal dye absorption capabilities were found to be associated with a low amount of Ti vacancies and high amount of N vacancies acting as donor states. The particles synthesized in water have the highest sorption capacity of dye achieving the value of 136.5 mg/g.

# 1. Introduction

Continuous population growth, overexploitation of water resources, uncontrolled urban water consumption, industrial pollution, intensification of agricultural production, global climatic changes, as well as natural limitations of freshwater resources, which account for only 2.5% of the world's total water volume, are the main reasons for the growing water crisis in the world.

Water affects all spheres of human life - from agricultural and industrial development to political and humanitarian processes. Water scarcity has a significant impact on agriculture, which is the main consumer of water resources (70% of all water withdrawals). Annual global industrial water use, which was 725 km³ in 1995, is expected to rise to 1,170 km³ by 2025, accounting for about 24% of total water use.

Another important problem, especially in developing countries, is providing the population with sewerage, i.e. disposal of domestic wastewater from the territory of settlements. At the beginning of the second decade of the XXI century, half of the world's population lives in cities, and by 2030 this figure will increase to 2/3 of the world's population. In social and medical aspects, not only the availability and quantity of drinking water, but also the quality of its regeneration and purification are important.

In recent years, the problem of wastewater has become increasingly acute and relevant throughout the world. In the process of economic activity, modern society consumes increasing amounts of water, most of which as a result becomes contaminated with a variety of substances. When they get into the environment, the ecology suffers enormous damage, and therefore they are subject to compulsory treatment. In order to provide it properly, it is necessary to use special equipment and technological complexes, with the help of which the established norms of wastewater pollution, defined in the relevant documents, are achieved. Anthropogenic factors of wastewater pollution are quite diverse and lead to the presence of mechanical, chemical and biological impurities, which are subject to removal by treatment facilities. As a rule, they are contained in wastewater in complex, in various concentrations, which significantly complicates the solution of the problem of wastewater treatment.

The introduction of modern technologies plays a significant role in the wastewater treatment process. For example, membrane technology can effectively remove pollutants as well as remove antibiotics from water. At the same time, the use of electrochemical processes in wastewater treatment and membrane distillation facilitates the recovery of water for reuse.

Wastewater treatment methods can be classified as mechanical, chemical, physical and biological methods. Mechanical method includes the use of mechanized grids, sedimentation of wastewater, treatment in hydrocyclones, use of sand traps, centrifugation, filtration and microfiltration. The chemical method involves oxidation and reduction, neutralization, sedimentation of suspended solids. The physical method uses magnetic and electromagnetic treatment, ultrasonic and ultraviolet treatment, ionizing radiation. Separately distinguish a combined method - physical and chemical treatment of wastewater. Here are applicable cleaning reagents - coagulation and flocculation, flotation and electroflotation, ion exchange, sorption, extraction, electrolysis. Biological and biochemical method is represented by anaerobic and aerobic treatment, wastewater disinfection, biological ponds. The final method is combined wastewater treatment - hyperfiltration, electrochemical wastewater treatment are used here.

One of the effective ways of cleaning from organic pollutants is photocatalytic method. The best known photocatalyst is titanium dioxide, but it is active only in the ultraviolet region of the spectrum. Because of this, it has not been industrially used so far. Organic pollutants are usually removed from wastewater using biological methods or sorbents. The biological method requires quite a lot of time for bacteria to work. And sorption method is quite fast and effective, but requires disposal of the used sorbent. Sorption purification is a process of absorption of pollutants by solid substances - sorbents. It allows to purify water from organic impurities, including those not removed by other methods. At present, a wide variety of sorbent materials are used: coal, expanded clay, zeolites.

The key point of the best available technologies is to minimize the use of hazardous and non-degradable substances in wastewater treatment. Within the framework of implementation it is necessary to refuse the use of chlorine and other hazardous reagents during production processes, to organize separate collection and disposal of residues of used substances, as well as to introduce environmentally safe reagents into production. In this regard, it is important to search for and develop effective reusable sorbents.

The use of porous nanomaterials as sorbents is a promising task for wastewater treatment. Nanomaterials and nanoparticles have a number of advantages: small size, high specific surface area, surface charge. Manufacturing of nanoparticles can be realized in different ways. One of the rather simple and reliable methods of nanoparticle synthesis is laser ablation in liquid, which has such advantages as absolutely clean surface, safe nanoparticles, uncompensated charge.

Nanoparticle systems (NPs) are popular research objects and potentially promising for applications in various branches of science and technology. They can be synthesized chemically or physically [1-5]. However, despite the prevalence of wet chemistry methods for synthesis of NPs, this approach typically uses many precursors in multi-step synthesis procedures, which negatively affects a critical parameter for catalytic applications - the purity of NPs surface. In addition, chemical methods often use complex multi-step procedures, and the resulting nanomaterials have low colloidal stability and must be further stabilized, which also significantly limits their possible applications. We supposebelieve that the most attractive solution in this context is the use of pulsed laser ablation in liquid (PLAL) for NPs synthesis [5-8]. Due to the nature of physical interaction of high-intensity laser radiation with matter, this method opens up the possibility for fabrication of colloidal solutions of ultrapure NPs with a ligand-free surface [5,9,10]. In addition, laser-ablated NPs typically demonstrate high colloidal stability associated with the natural uncompensated surface charge due to the extremely nonequilibrium processes of nanocluster formation during the synthesis process. This unique feature can be also effectively used in removal of charged or polar contaminants from aquatic environments. Moreover, the method of pulsed laser ablation in liquids is scalable and highly efficient. Aswhile the productivity of PLAL technique exceeds 550 mg/h, laser ablation becomes a more cost-effective approach compared to chemical methods [11]. Simplicity of implementation, flexibility and the ability to synthesize NPs from almost any inorganic material brings additional attractiveness for a wide range of applications, including catalytic and sorption for water purification.

Titanium nitride may be one of the promising materials as sorbents. Titanium nitride nanoparticles (TiN NPs) have photoluminescent properties in the visible range [12] It is shown that TiN nanoparticles synthesized by laser ablation have a strong phototherapeutic effect at 750-800 nm excitation [13]. The results obtained indicate the high safety of laser synthesized TiN NPs for biological systems, which promises a serious development of phototherapeutic modalities based on them. The method of laser ablation in liquid, taking into account the selection of laser operation modes and the type of solvent, allows to vary the size and shape of synthesized nanoparticles. Recently, TiN is considered as a promising material for biomedical applications [13]. Cube-shaped TiN nanoparticles were synthesized by laser ablation method in a gas ablation chamber, and their optical and colorimetric properties were determined [14]. Although pulsed-laser ablation in gaseous media is a widely-spread method, it requires a vacuum environment to ensure the pristinity of the obtained NPs. The necessity to operate vacuum chamber and associated pumping setup increases time consumption and decreases cost-efficiency associated with the NPs manufacturing. These issues are to some extent alleviated in PLALlaser ablation in liquid (LAL) technique. Previously, our team used PLAL to obtain titanium nitride NPs and considerably tune their oxidation state by alternating the ablation media [15]. We have also implemented two-step laser treatment to create ultrafine biocompatible NPs capable of photothermal biomedical treatment [16].

In this work, we investigate the possibility of water purification via dye absorption by the titanium nitride nanoparticles produced by laser ablation. Cationic dyes pose a severe threat to aquatic life, being toxic and carcinogenic bioactive agents [17]. In addition, dye-loaded water is often used as a model environment of contaminated aqueous media, thus allowing for the robust assessment of the wastewater treatment techniques aiming at the sustainable and cost-effective purification of potable water. A plethora of materials have been tested as dye removal sorbents, namely, silver-nanoparticle-based composites [17], functionalized nanotubes [18], graphene oxide [19]. However, the application of silver and sp2-carbon-based materials can lead to the release of metal ions or graphene/graphene oxide flakes into the liquid, which may pose a threat in itself due to the bioactivity of above-mentioned materials. Thus, safe and biocompatible alternatives to conventional adsorbents are required to be investigated to make water purification technology benign and harmless. Titanium nitride nanoparticles produced by laser ablation in liquid are ideal candidates for such investigation, owing to their high surface to volume ratio and tunable stoichiometry. Stoichiometry variation can provide a platform to control the type of the defects acting as cationic-dyes-selective adsorption sites.

In our study, we investigate the interplay between the ablation media, stoichiometry of the nitrogen-deficient with specified properties, namely, potentially high adsorption properties.

## 2. Materials and Methods

### 2.1. Laser-ablative synthesis of TiN NPs

Colloidal solutions of TiN nanoparticles were synthesized by a method of pulsed laser ablation in liquids, similar to our previous works [13, 15, 16]. Schematic representation of the laser ablation process is shown in Figure 1. A Ø3 mm beam from Yb:KGW system (1030 nm, 250 fs, 25 µJ, 200 kHz, TETA-20 model, Avesta, Russia) was used as a source of laser radiation. The bulk crystalline target of TiN (99.999% purity, Girmet, Russia) was fixed vertically inside a glass chamber (BK-7, wall thickness 3 mm) filled with 20 ml of liquid media (deionized water, acetone, acetonitrile). Laser beam was focused on the target surface by F-Theta lens (100 mm focal distance, Thorlabs, USA). To improve the synthesis productivity the liquid thickness between the target and chamber wall was minimized down to 2.5 mm. To avoid ablation from one spot the laser beam was continuously moved over a Ø20 mm area on the target surface with scanning geometry like self-closed helix, with 5 m/s speed by galvanometric scanner (LScan-10, Ateko-TM, Russia). The duration of laser ablation was 30 minutes. The NPs formation resulted in deep-blue coloration of the colloidal solution.

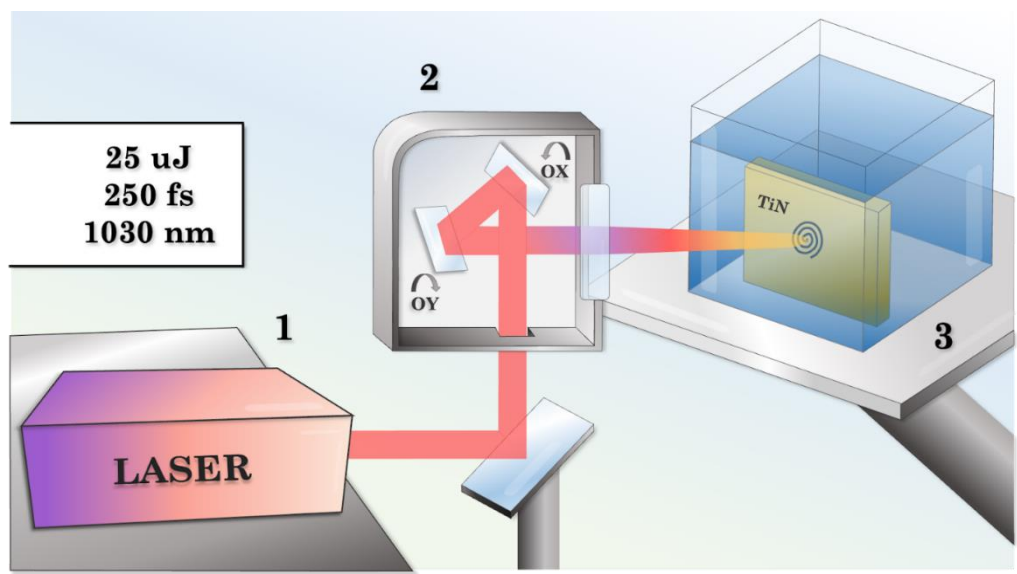

**Figure 1.** Schematic representation of experimental setup for laser ablation in liquids. (1) – fs Yb:KGW laser, (2) – galvanometric scanner with F-theta lens, (3) – ablation chamber with TiN target and prepared colloidal solution.

*2.2. Characterization of TiN NPs*

The morphology, chemical composition and size characteristics of the TiN NPs were characterized by means of scanning electron microscopy (SEM) system (MAIA 3, Tescan, Czech Republic) coupled with EDX detector (X-act, Oxford Instruments, UK) and using a JEOL JEM-2100 transmission electron microscope (Japan) at an accelerating voltage of 200 kV. The JEOL JEM-2100 transmission electron microscope is equipped with an Aztech X-Max 100 energy dispersive analysis attachment. SEM images and EDX analysis were obtained at 30 keV accelerating voltage. Samples for the SEM and EDX measurements were prepared by dropping 1-5 μL of colloidal solution onto a cleaned monocrystalline Si substrate and subsequent drying at ambient conditions.

The size distribution of the synthesized NPs were obtained by analysis of the SEM images in the ImageJ software environment with circle fit approximation. The final distribution was based on measurement of 300-500 NPs diameter.

The distribution of hydrodynamic diameter and ζ-potential were measured by the dynamic light scattering technique using a Zetasizer Nano ZS device (Malvern Instruments, Malvern, UK). The mode values ± half-width of the peak of number-weighted size distributions were used for analysis. The Smoluchowski approximation was used for the ζ-potential calculation.

The optical extinction spectra were analyzed by a spectrophotometry method with MC 122 system (SOL Instruments, Belarus) in the spectral range of 330–1000 nm. For the measurement were used glass cuvettes with 10 mm optical path length filled with 1 ml colloidal solution.

Raman spectra were acquired with a Horiba LabRAM HR Evolution (HORIBA Ltd., Kyoto, Japan) confocal Raman microscope. The excitation wavelength was 633 nm. 100× /N.A. = 0.90 microscope objective and 600 lines/mm diffraction grating were used. Spectra were obtained from the NPs deposited on cover glasses by drop-casting. Incident laser power of 0.1 mW was used for the studies to avoid sample degradation. Acquisition time of each point was 30×10 sec. Five spectra were obtained for each sample, and the most typical one was analyzed. Only a slight variation of the spectra in various points was observed.OD1.6 filter of incident laser power (4.5 mW) was used for the studies.

*2.3. Sorption properties*

TiN NPs were transferred to water after synthesis in various solvents. The NPs were centrifuged for 20 minutes at 14000 RPM to sediment the TiN NPs. The solutions atop the sedimented TiN NPs were removed, and the sediment was resuspended in water. This process was repeated 2 times to minimize the presence of organic solvents. After the described procedure, the concentration of TiN NPs were 0.1 g/L.

We evaluated the sorption ability of Titanium Nitride nanoparticles using a Methylene Blue (MB) solution in batch sorption experiments. The TiN NPs were synthesized in different organic solutions before being transferred to DI water. For each test, the TiNNPs were combined in a 2 ml Eppendorf tube with 1 ml of a 20 mg/L MB solution at room temperature (25 ± 2 °C). The sample solutions, along with their blanks and experimental controls (which had no sorbent), were stirred for 2 minutes using a magnetic stirrer. After stirring, we used an UV-visible absorption spectrophotometer (Cary 5000; Agilent Technologies) to measure the absorption spectra of the filtrate, which was obtained by separating the TiN NPs using a centrifuge (3 min, 14000 RPM). The amount of adsorbed methylene blue was determined by the difference between the initial concentration of MB and the remaining dye in the solution, which was calculated by the peak intensity of 664 nm adsorption band using a preliminary calibration curve.

## 3. Results and Discussion

### 3.1. laser synthesis of nanoparticles

To conduct experiments on dark adsorption of methylene blue, three different types of TiN NPs were synthesized: in water, acetone and acetonitrile. The TiN NPs obtained in acetone and acetonitrile were then transferred into an aqueous solution for further tests. All synthesized nanostructures had a spherical morphology, but the mode and half-width of the size distribution were slightly different (Figure 2 (a,b,c)). Laser ablation of a crystalline TiN target in water (TiN-$H_2O$) resulted in the formation of hollow nanostructures with cavities inside, which is typical for laser-synthesized TiN NPs (Figure 2a, Supplementary Figure S1) [16]. Moreover, such NPs had a size distribution mode of 60–70 nm and a FWHM of about 60 nm (Figure 2(a)). On the other hand, laser ablation in acetone and subsequent transferring to water (TiN-AC) resulted in the formation of intact TiN NPs with a distribution mode of 50–60 nm and a relatively narrow FWHM of 38 nm (Figure 2(b)). The TiN NPs synthesized in acetonitrile and transferred to water (TiN-AN) demonstrated the lowest size distribution mode of 40–50 nm and FWHM of about 30 nm (Figure 2(c)). The obtained results from the analysis of SEM images of different TiN NPs are in high correlation with the analysis of colloidal solutions by the DLS method (Figure 2d).

Despite differences in size characteristics all the resulting TiN NPs had optical extinction spectra typical for group IV transition metal nitrides (Figure 2(e))[20]. The spectral shift of the extinction peak, as well as the change in its intensity within the synthesis in various liquid media is associated with increased oxidation of the surface of TiN NPs during laser ablation in water and also TiN NPs size characteristics [15]. Moreover, the increased oxygen content in the water sample is also confirmed by EDX analysis (Figure 2(f)). In addition, laser ablation of TiN in an organic solvent such as acetone or acetonitrile could lead to the formation of complex carbides on the surface of TiN NPs, which in this case is indirectly confirmed by the increased carbon content [15].

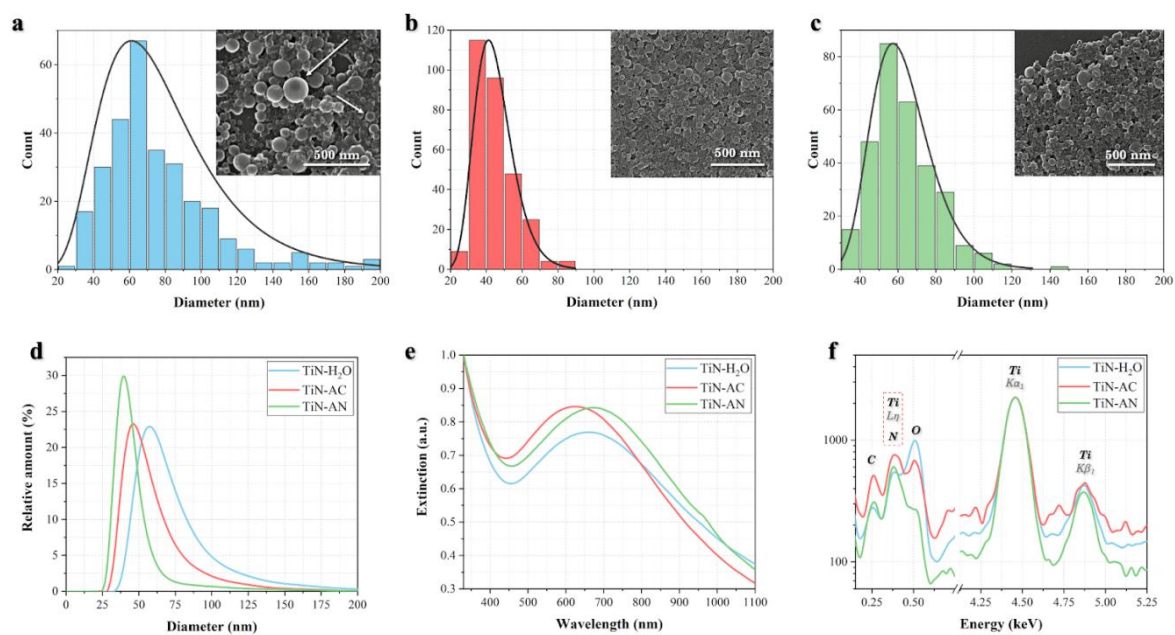

**Figure 2.** Size distribution and SEM image of TiN NPs synthesized in (a) water, and transferred to water from (b) acetone or (c) acetonitrile. In the insets of Figs. 2(a-c), SEM images of obtained materials are presented. Comparison of (d) hydrodynamic size distribution, (e) extinction spectra, and (f) EDX spectra of different types of TiN NPs.

Figure 3(a) shows a typical view of TiN nanoparticles in TEM. Selected-area electron diffraction (SAED) pattern presented in the inset of Fig. 3 demonstrates a set of point reflexes, which proves the crystallinity of obtained NPs. For cubic structures such as TiN, interplanar spacing $d_{hkl}$ relation to the lattice parameter a is represented by the following formula, where h, k and l are Miller indices:

$$d_{hkl} = a/\sqrt{h^2 + k^2 + l^2}$$

For the investigated structure, we observe the SAED-derived interplanar spacings at 0.25, 0.21, 0,15, 0.13 and 0.10 nm. Previously, interplanar spacings of 0.24 nm, 0.21 nm and 0.14 nm were identified as (111), (200) and (220) planes of nitrogen-deficient TiN [21]. Imperfect alignment between the theoretical and experimental values indicates that the lattice parameter of obtained NPs is not in a good correspondence with literature data. To identify the lattice spacing, we assumed that 0.13 nm reflex is related to the cumulative contribution of closely-positioned (311) and (222) peaks, while 0.10 nm reflex is related to the reflex from (331) and (420) peaks (see Fig. 3(b). Previously, the manifestation of these reflexes in diffraction patterns was reported in [22]. Basing on these assumptions, we have calculated the lattice constant of 0.4345 nm considerably differing from the 0.4236 nm of pristine TiN[23]. Previously, ~0.43 nm lattice parameter was reported for ion-irradiated TiN [24], thus confirming heavily defected and distorted structure of obtained NPs.

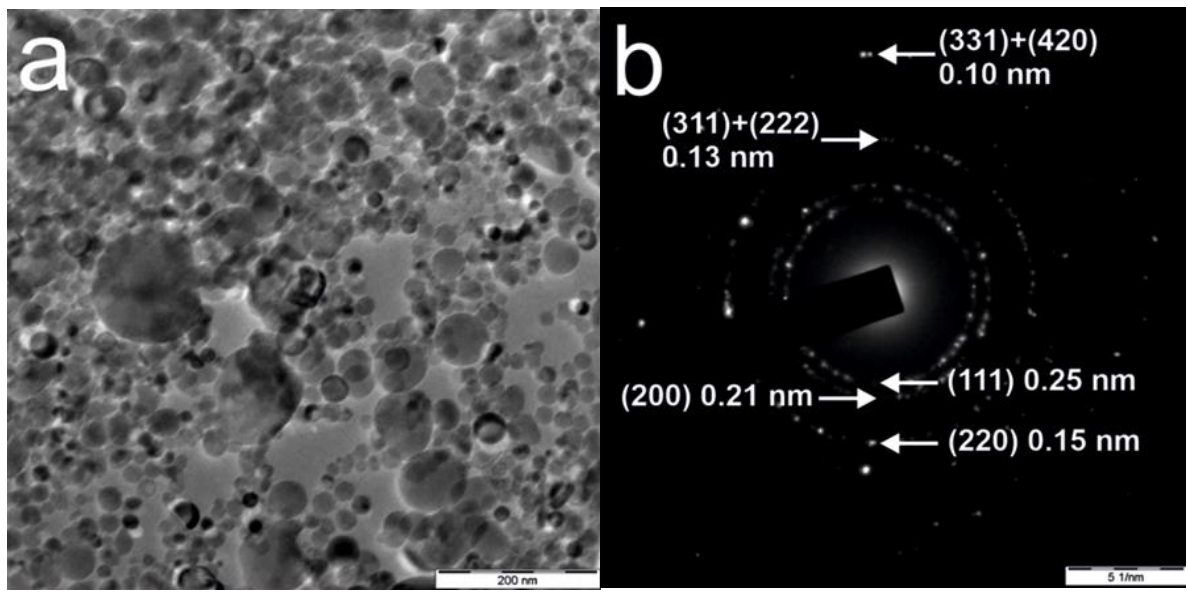

**Figure 3.** (a) TiN particles synthesized in acetone and transferred to water ("acetone to water"). (b) SAED pattern.

Figure 4 shows the data on the quantitative content of elements in TiN nanoparticles. Quantitative analysis of elements was performed from nanoparticles (Spectrum 2, Figure 4). For this purpose, a solution of titanium nitride nanoparticles "acetone to water" was dripped onto a copper grid covered with a carbon film and measured in a transmission microscope in the STEM EDX mode. The aluminum content is within the margin of error, and its presence can be disregarded. There is also some oxygen content associated with oxidation during the transfer of TiN nanoparticles into water. Fig. 4 (c,d) shows that the distribution of titanium and nitrogen in the nanoparticles is relatively uniform and it can be concluded that the laser synthesis of TiN nanoparticles is quite good. The titanium content prevalence over nitrogen allows us to assume that obtained particles are under-stoichiometric. However, EDX results tend to vary considerably from area to area, thus not allowing for a reliable determination of the stoichiometric ratio of the sample. To address this issue, we have carried out Raman studies presented in Section 3.2.

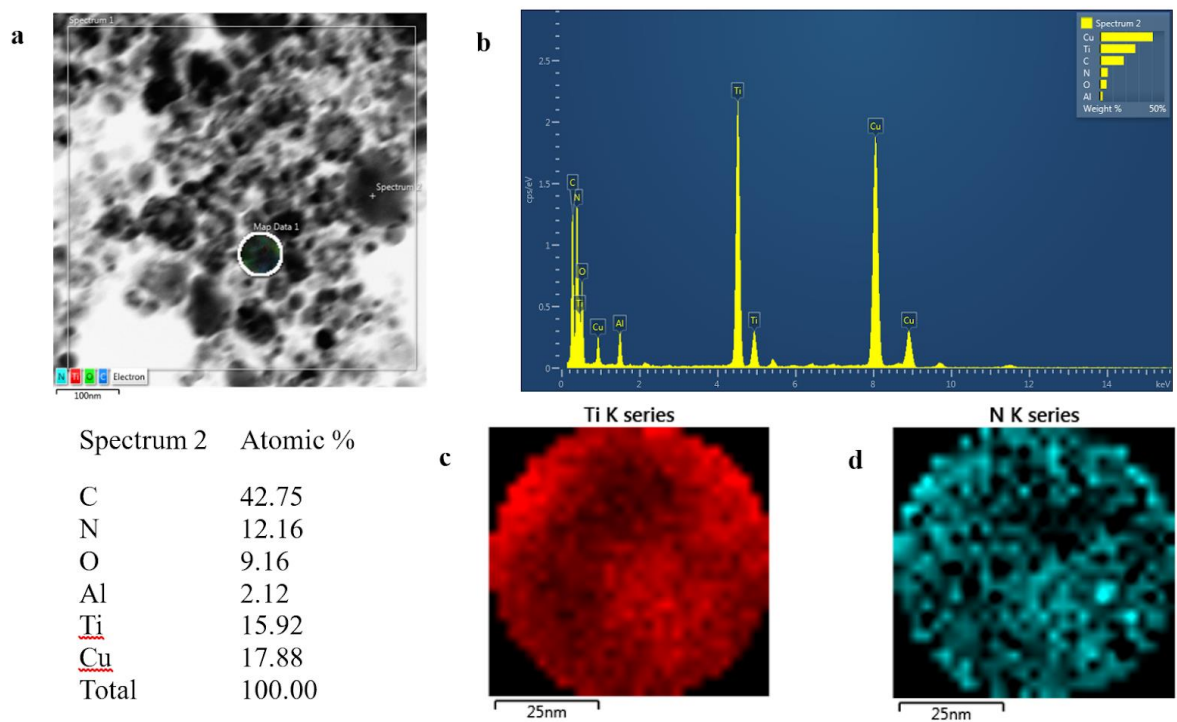

**Figure 4.** TEM EDX analysis of TiN nanoparticles synthesized in acetone and transferred to water.

*3.2. Zeta potential, adsorption and removal of cationic dyes*

The obtained colloidal solutions demonstrated high colloidal stability, as confirmed by the Zeta-potential measurements (Figure 5). All obtained NPs had a strictly negative Zeta-potential. However, the value of Zeta-potential and colloidal stability are affected not only by the type of NPs material, but also by the liquid in which the synthesis was carried out [25]. TiN NPs obtained in water showed the highest potential of (-48 ± 25) mV, while for the acetonitrile (-28 ± 13) mV and acetone sample (-15 ± 18) mV. The presence of a relatively high uncompensated charge on the surface of NPs is one of the many unique features of the laser ablation in liquid.

In addition to increased colloidal stability without the extra steps of TiN NPs surface modification, a high negative Zeta-potential can also act as an additional attractive mechanism for the electrostatic adsorption of a pollutant in solution. To confirm this hypothesis, we measured the amount of adsorbed methylene blue dye across various concentrations of nanoparticles. The obtained dependencies reveal that the particles synthesized in water have the highest sorption capacity at 136.5 mg/g. In contrast, particles synthesized in acetonitrile and acetone and then transferred to water demonstrate significantly lower capacities at 52.8 and 22.5 mg/g, respectively. These results are well aligned with the Zeta-potential, reinforcing the electrostatic nature of methylene blue adsorption.

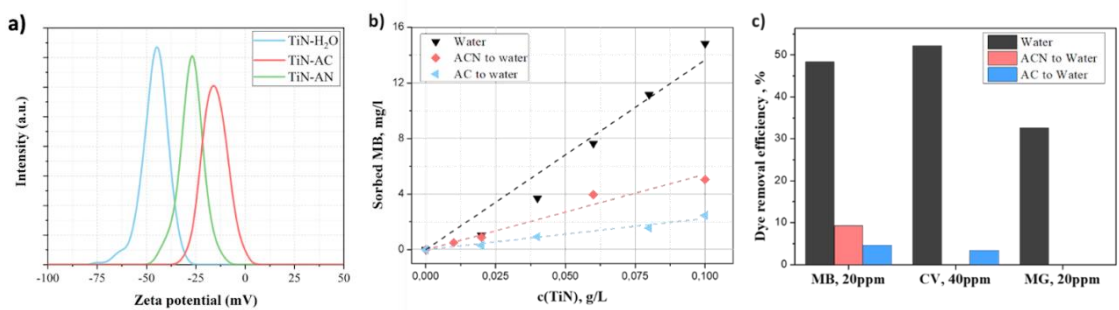

**Figure 5.** (a) Zeta potential distribution (a), adsorption isotherm Methylene blue (b), cationic dye removal efficiencies (c) of different types of TiN NPs.

Given the electrostatic nature of Methylene Blue (MB) adsorption, we found it essential to examine the sorption efficiency of other cationic dyes, such as Crystal Violet (CV) and Malachite Green (MG). The nanoparticles were synthesized in water effectively removed all these dyes from the solutions. Notably, TiN NPs in water showed the highest adsorption efficiency Crystal Violet due to tricationic nature dye. The lack of adsorption of CV and MG on TiN NPs synthesized in other solutions is due to steric hindrance, as these compounds have a larger molecular size compared to MB. This theory is supported by the presence of cavities in nanoparticles synthesized in water (Supplementary Figure S1), which provides a significant adsorption capacity for all the dyes we studied (Table 1).

Table 1. Structural and physicochemical properties of the TiN NPs based sorbents.

| Type NP | Hydrodynamic size (nm) | Zetapotential (mV) | Methylene blue adsorption capacity (mg/g) | Dyeremovalefficiency, % * | | |
|---|---|---|---|---|---|---|
| | | | | Methyleneblue, (20ppm) | Crystal violet (40ppm) | malachitegreen (20 ppm) |
| Water | 60-70 (+-60) | -48 | 136.5 | 48 | 52 | 32 |
| Acetonitriletowater | 50–60 (+-38) | -28 | 52.8 | 9 | 1 | 0 |
| Acetonetowater | 40-50 (+-30) | -15 | 22.5 | 4 | 3 | 0 |

* concentration of TiN NPs 1 g/L

*3.3. Raman studies*

Raman spectra of ablated targets and all of the investigated samples show three peaks at 210, 310-330 and 550 cm$^{-1}$ (see Fig. 6(a,b)). For TiN face-centered cubic lattice with $O_h$ full octahedral symmetry, Raman modes are forbidden by the selection rules [26]. Thus, the manifestation of Raman peaks in the spectra of TiN-based samples is related either to the symmetry breaking by defects or to the formation of non-Ti-N bonding. Emergence of the peaks at 210-220, 310-330 and 540-560, 610-640 cm$^{-1}$ with relatively intensive low-wavenumber peaks is related to the formation of under-stoichiometric TiN$_{1-x}$ structure [27]. In Raman spectra of TiN$_{1-x}$, acoustic phonon bands (210-220 cm$^{-1}$ for transverse acoustic phonons (TA) and 310-330 cm$^{-1}$ for longitudinal acoustic phonons (LA)) are related to the presence of nitrogen vacancies, while optical-phonon-related lines (540-560 cm$^{-1}$ for transverse optical phonons and 610-640 cm$^{-1}$ for longitudinal optical phonons (TO/LO)) are ascribed to titanium vacancies [28, 29]. Presence of all four bands in the investigated spectra shows that analyzed structures possess both Ti and N vacancies.

As Raman scattering intensity of TiN$_{1-x}$ is directly related to the defects concentration, Raman spectroscopy provides a handy tool to assess the variation of the titanium nitride target surface. As shown in Fig. 6(a), no additional lines were observed after target irradiation, thus confirming that laser ablation hasn't considerably changed the materials' structure. However, line intensities, especially the ones of acoustic modes, considerably rise after the irradiation, thus showing laser-induced defect introduction into the target surface resulting in prominent formation of nitrogen vacancies.

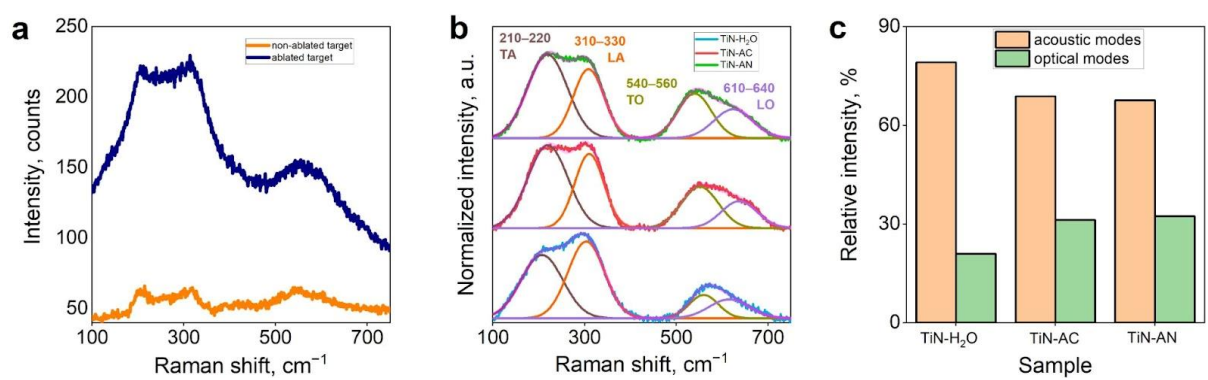

**Figure 6.** (a) Non-normalized Raman spectra of non-ablated and ablated titanium nitride targets. (b) Normalized Raman spectra of the TiN$_{1-x}$ NPs obtained in various conditions. Positions and origins of the bands (transverse/longitudinal acoustic/optical phonons) are indicated. (c) Relative intensities of acoustic and optical bands derived from the fitting of background-subtracted spectra.

To process the spectra of the resulting powders in a comprehensive way, we have subtracted their baseline by OriginPro software, normalized and fitted them with four Gaussian lines positioned in the ranges typical for the wavenumbers TiN$_{1-x}$ phonons by MagicPlot software. One distinction to be addressed is a relatively low intensity of the LO/TO phonon bands for the NPs produced by ablation in water. Quantitative assessment of this feature is presented in Fig. 6(c) which shows relative areas filled by acoustical and optical modes. In line with the discussion above, we can deduce that spectra fitting shows a relatively low amount of titanium vacancies and/or a relatively high concentration of nitrogen vacancies being present in the particles ablated in water.

*3.4. Discussion*

In pristine TiN, titanium atoms possess +3 oxidation state, while nitrogen atoms have -3 oxidation state. Therefore, nitrogen-deficient TiN$_{1-x}$ nanoparticles obtained by ablation have a considerable amount of donor defects emerging from uncompensated Ti electrons, which, as shown by Zeta-potential studies presented in Section 3.2., leads to their negative surface potential of NPs. For reduced graphene oxide, it was shown that negative Zeta-potential leads to its electrostatic absorbance of cationic dyes [30]. In our case, however, a similar effect is caused by vacancies rather than surface groups. Notably, Raman studies show that water-ablated nanoparticles are comparatively Ti-rich or N-deficient. In terms of the electrostatic aspects, both titanium abundance and nitrogen scarcity will lead to the affluence of donor states, thus increasing the absolute value of negative zeta potential. As indicated by Section 3.2, Ti-richness/N-deficiency of the TiN$_{1-x}$ particles produced in water considerably improves their cationic dyes uptake, thus making them promising candidates for the universal absorbers of cationic dyes. The schematic summary of our findings is presented in Fig. 7.

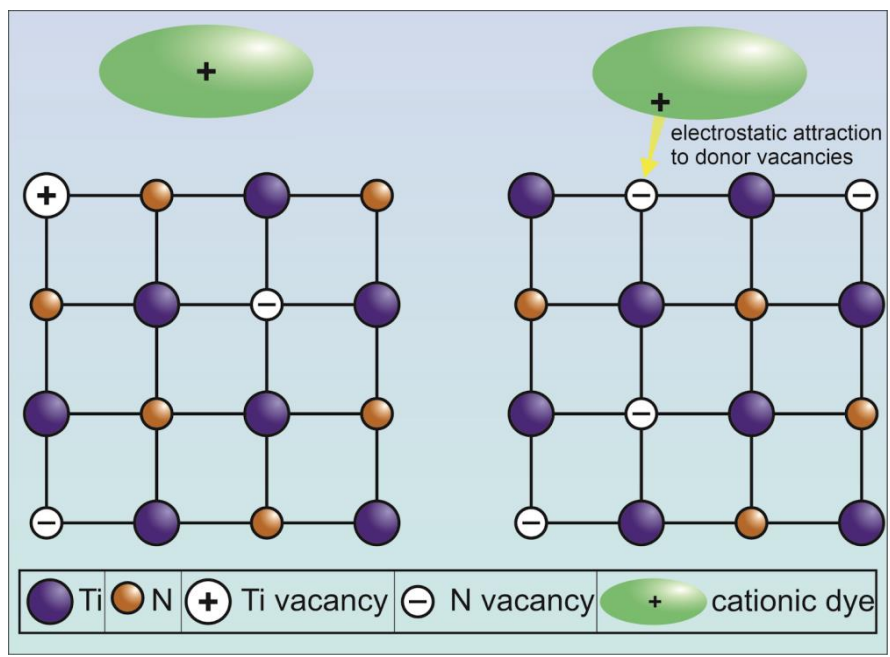

**Figure 7.** Representation of relatively moderate absorbance of cationic dyes by TiN$_{1-x}$ (left side) and superior absorbance by donor-defect-rich TiN$_{1-x}$.

We should also note that absorbance spectra of TiN$_{1-x}$ NPs produced by ablation in water demonstrate the lineshape which is untypical for pristine CV, which shows that chemical processes take place throughout the interaction of dye and water-ablated NPs (Figure S3). This effect requires further investigation. Notably, its manifestation is in line with our findings regarding donor states abundance in water-ablated TiN$_{1-x}$, as presence of electron donors can enhance the photocatalytic activity of the samples leading to the dye degradation induced by reactive oxygen species [31].

**4. Conclusions**

The adsorption properties of titanium nitride nanoparticles synthesized in water, acetone and acetonitrile, subsequently transferred to water, have been studied. As a result of laser ablation in liquid, titanium nitride nanoparticles are titanium-rich and nitrogen-deficient, leading to an abundance of donor states and increasing the absolute value of the negative zeta potential. The Ti-rich/N-deficient TiN$_{1-x}$ particles produced in water significantly improve their uptake of cationic dyes, making them promising candidates for cationic dye uptake. The particles synthesized in water have the highest sorption capacity of 136.5 mg/g. The high adsorption properties of TiN NPs has an electrostatic nature. Note that the high adsorption by titanium nitrides nanoparticles is in good agreement with the conclusions regarding the abundance of donor states in ablated TiN$_{1-x}$ in water, since the presence of electron donors can enhance the photocatalytic activity of the samples, leading to dye degradation under the action of reactive oxygen species.

Note that TiN nanoparticles can be reused after adsorption. It is necessary to anneal the nanoparticles at the combustion temperature of organic impurities. For most organic pollutants, annealing at 300 C for 30 minutes is sufficient.

**Supplementary Materials:** Figure S1: TEM images of TiN NPs synthesized in water, acetone, and acetonitrile; Figure S2: The absorption spectra of methylene blue at different concentrations (a), and the calibration curve for absorbance to methylene blue concentration (b); Figure S3: Uv-vis absorbance spectra of initial cationic dyes (a) Methylene Blue (MB), (b) Crystal Violet (CV) (c) Malachite Green (MG) and after addition TiN NPs obtained in various conditions.

**Author Contributions:** Conceptualization, A.V.K., Q.S. and V.S.V.; methodology, X.Z., X.Y., A.A.P.; validation, S.M.N., A.V.A., G.I.T.; formal analysis, S.M.N., A.V.A., G.I.T.; investigation, A.V.S., I.V.M., I.A.Z., G.V.T., D.V.D., D.I.T., M.S.S., I.V.S., S.M.K.; writing—original draft preparation, A.V.S., G.V.T., I.V.M., I.A.Z.; writing—review and editing, A.V.S., A.V.K., V.S.V.; visualization, D.I.T., M.S.S., I.V.S., S.M.K.; supervision, A.V.K., Q.S. and V.S.V.; project administration, A.V.A., X.Z.

**Funding:** This research was funded by RSCF, grant number 22-19-00094 and supported by the Ministry of Science and Higher Education of the Russian Federation (FSMG-2024-0014).

**SupplementaryMaterials:**

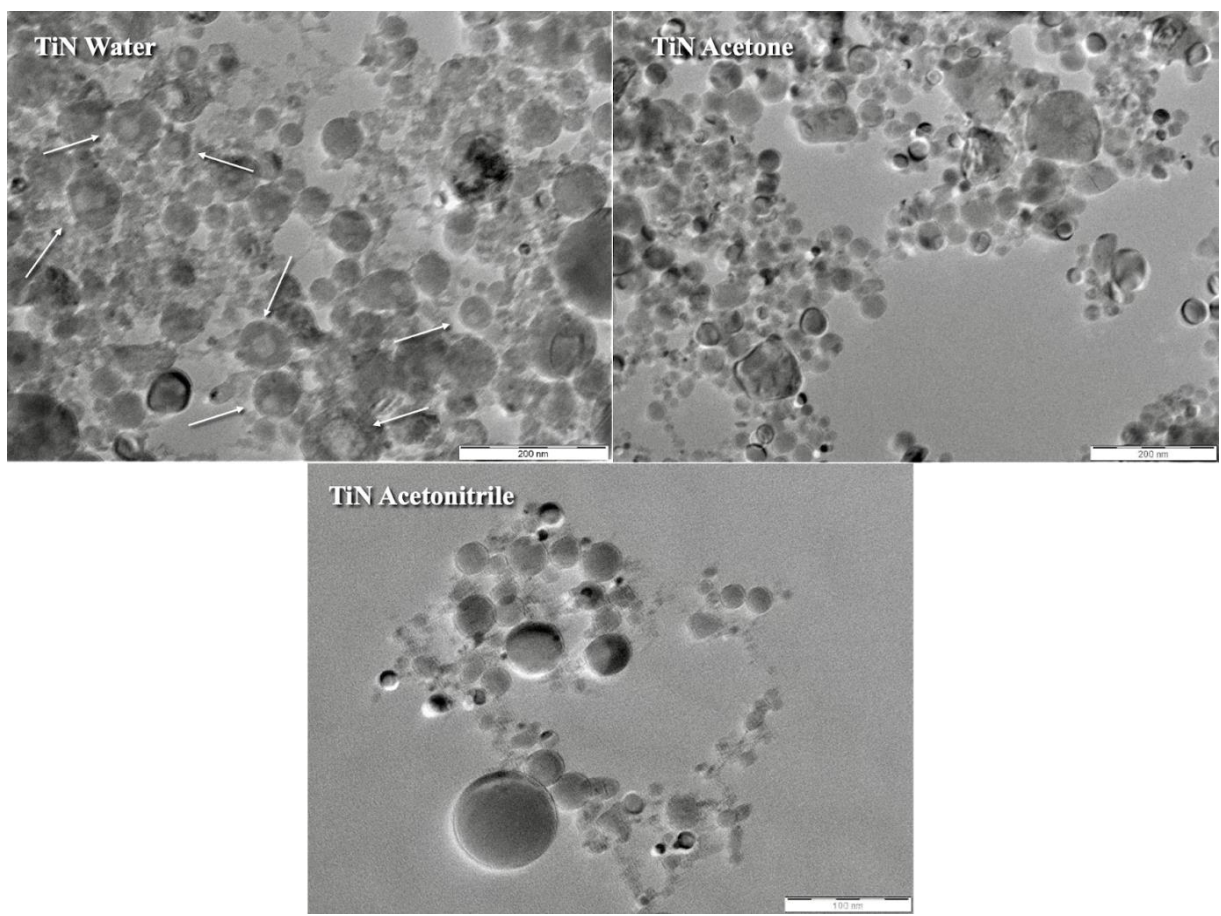

Figure S1. TEM images of TiN NPs synthesized in water, acetone, and acetonitrile.

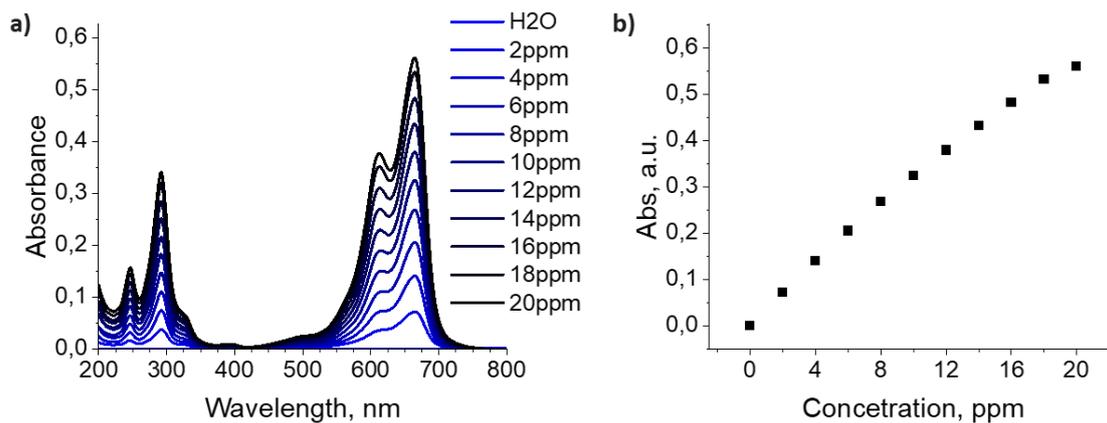

Figure S2. The absorption spectra of methylene blue at different concentrations (a), and the calibration curve for absorbance to methylene blue concentration (b).

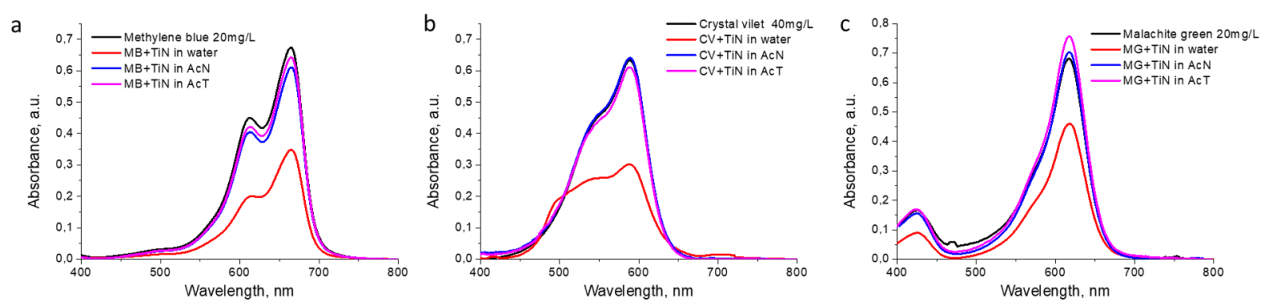

Figure S3. Uv-vis absorbance spectra of initial cationic dyes (a) Methylene Blue (MB), (b) Crystal Violet (CV) (c) Malachite Green (MG) and after addition TiN NPs obtained in various conditions.